\documentclass[12pt]{article}

\def\eg{{\it e.g.}\ }

\def\be{\begin{equation}}
\def\ee{\end{equation}}
\def\bB{{\bf b}}
\def\bv{{\bf v}}

\begin{document}
\begin{center}
{\bf On the von K\'arm\'an-Howarth equations for Hall MHD flows}\\

S\'ebastien Galtier\\

Institut d'Astrophysique Spatiale (IAS), B\^atiment 121, F-91405 Orsay (France);
Universit\'e Paris-Sud XI and CNRS (UMR 8617)\\
\end{center}
\vskip0.5truein

\begin{abstract}
The von K\'arm\'an-Howarth equations are derived for three-dimensional (3D) Hall magnetohydrodynamics 
(MHD) in the case of an homogeneous and isotropic turbulence. From these equations, we derive 
exact scaling laws for the third-order correlation tensors. We show how these relations are compatible 
with previous heuristic and numerical results. These multi-scale  laws provide a relevant tool to 
investigate the non-linear nature of the high frequency magnetic field fluctuations in the solar wind 
or, more generally, in any plasma where the Hall effect is important. 
\end{abstract}

Turbulence remains one of the last great unsolved problem in classical physics which has evaded 
physical understanding and systematic description for many decades. For that reason, any exact 
results appear almost as a miracle. In his third 1941 turbulence paper, Kolmogorov found that an 
exact and nontrivial relation may be derived from Navier-Stokes equations -- which can be seen as 
the archetype equations for describing turbulence -- for the third-order longitudinal structure function 
(Kolmogorov, 1941). Because of the rarity of such results, the Kolmogorov's four-fifths law is considered as 
one of the most important results in turbulence (Frisch, 1995). 

The derivation of the Kolmogorov's law uses earlier exact results found by von K\'arm\'an and 
Howarth in 1938 (von K\'arm\'an and Howarth, 1938): it is the well-known von K\'arm\'an-Howarth (vKH) 
equation that 
describes the dynamical evolution of the second-order correlation tensors. Very few extensions of 
such results (vKH equations and four-fifths law) to other fluids have been made; it concerns scalar 
passively advected (Yaglom, 1949), such as the temperature or a pollutant in the atmosphere, and 
astrophysical magnetized fluid described in the framework of MHD 
(Chandrasekhar, 1950; Politano and Pouquet, 1998; Politano, Gomez and Pouquet, 2003). The 
addition in the analysis of the magnetic field and its coupling with the velocity field renders the 
problem more difficult and, in practice, we are dealing with a couple of equations. 

Signatures of turbulence in astrophysical flows are found in the solar wind (Goldstein and Robert, 1999;
Matthaeus et al., 2005), the interstellar (Elmegreen and Scalo, 2004ab), 
galactic and even intergalactic media (Govoni et al., 2006). In the case of 
the interplanetary medium, we have access to very precise {\it in situ} measurements which show, 
in particular, the existence of a steepening of the magnetic field fluctuations spectrum at frequencies 
higher than $1$Hz (Bale et al., 2005; Smith et al., 2006) whose origin may be attributed to nonlinear 
Hall-MHD processes (Galtier, 2006ab; Galtier and Buchlin, 2007). 
Efforts from observers are currently made to show the presence of intermittency at high 
frequency. In this quest, any theoretical or model predictions about moderate or high-order 
correlation tensors is particularly important for the understanding of solar wind, and more generally 
speaking, astrophysical turbulence. 

In this Letter, we derive the vKH equations for 3D Hall-MHD fluids. From these exact results, we 
give the equivalent of the Kolmogorov's four-fifths law for the velocity, magnetic and current density 
field correlations.

We start our analysis with the following 3D incompressible Hall MHD equations
\begin{eqnarray}
(\partial_t + \bv \cdot \nabla) \bv &=& - {\bf \nabla} P_* + \bB \cdot \nabla \, \bB
+ \nu \Delta \bv \, , \label{hmhd1} \\
(\partial_t + \bv \cdot \nabla ) \bB &=& \bB \cdot \nabla \, \bv 
- d_I \nabla \times [ (\nabla \times \bB) \times \bB ] + \eta {\Delta} \bB , \hspace{.4cm} \label{hmhd2}
\end{eqnarray}
with $\nabla \cdot \bv = 0$, $\nabla \cdot \bB = 0$. 
The magnetic field $\bB$ is normalized to a velocity ($\bB \to \sqrt{\mu_0 n m_i} \, \bB$, with 
$m_i$ the ion mass and $n$ the electron density), $\bv$ is the plasma flow velocity, $P_*$ is the total 
(magnetic plus kinetic) pressure, $\nu$ is the viscosity, $\eta$ is the magnetic diffusivity and $d_I$ is 
the ion inertial length ($d_I = c / \omega_{pi}$, where $c$ is the speed of light and $\omega_{pi}$ is the 
ion plasma frequency). 
Equations (\ref{hmhd1})--(\ref{hmhd2}) may be rewritten more compactly as 
\begin{eqnarray}
\partial_t v_i &=& - \partial_i P_* + b_\ell \partial_\ell b_i - v_\ell \partial_\ell v_i 
+ \nu \partial^2_{\ell \ell} v_i , \label{E1} \\
\partial_t b_i &=& b_\ell \partial_\ell v_i - v_\ell \partial_\ell b_i 
+ d_I (J_\ell \partial_\ell b_i - b_\ell \partial_\ell J_i) + \eta \partial^2_{\ell \ell} b_i , \hspace{.5cm} \label{E2}
\end{eqnarray}
where ${\bf J} = \nabla \times \bB$ is the normalized current density. Note the use of the Einstein's notation. 
We see immediately that the third-order tensors that will appear in our analysis will be a combination of the 
velocity, the magnetic field and the current density. This makes an important difference with Navier-Stokes 
fluids for which the tensor used to derive the $4/5$-law is built with the same (velocity) field. As shown 
below, it will have a direct impact on the kinematics.

Before deriving the vHK relations for Hall-MHD, one needs to introduce the kinematics adapted 
to this problem. The second-order correlation tensors, in the full isotropic and homogeneous case, 
may be written as (Batchelor, 1953)
\begin{eqnarray}
R^X_{ij} ({\bf r}) &=& \langle X_i ({\bf x}) X_j ({\bf x}^{\prime}) \rangle = F^X r_i r_j + G^X \delta_{ij} \, , 
\end{eqnarray}
where ${\bf x}^{\prime}={\bf x}+{\bf r}$ and $X=v$ or $b$. $F^X$ and $G^X$ are four arbitrary functions of 
$r^2$ which will be specified later. The divergence free condition, $\partial_{r_j} R^{X}_{ij} ({\bf r}) = 0$, 
on the velocity and the magnetic field leads to the relations
\be
4F^{X} + r \partial_r F^{X} + r^{-1} \partial_r G^{X} = 0 \, ,
\label{r1}
\ee
which will be used later. We now introduce the longitudinal and lateral functions as, respectively, 
\be
R^X_{\parallel \parallel} = X^2 f^X(r) \, \, \rm{and} \, \, 
R^X_{\perp \perp} = X^2 g^X(r) \, . \label{r1c} 
\ee
The reference direction is the vector separation ${\bf r}$ such that, for example, the parallel component 
is the one along ${\bf r}$. The correlation functions $f^{v,b}$ and $g^{v,b}$ are mainly decreasing 
(see \cite{batch} for more details about the function $g$) and satisfy the condition 
$f^{v,b}(0)=g^{v,b}(0)=1$. From relations (\ref{r1})--(\ref{r1c}), we obtain 
\be
R^X_{ij} ({\bf r}) = 
X^2 \left(f^X \delta_{ij} + ({r \over 2} \delta_{ij} - {r_i r_j \over 2r}) \partial_r f^X \right) \, . 
\ee

The third-order correlation tensors that will appear in our derivation are (Batchelor, 1953)
\begin{eqnarray}
S^1_{ijk} ({\bf r}) &=& \langle v_i ({\bf x}) v_j ({\bf x}) v_k ({\bf x}^{\prime}) \rangle \\
&=& A_1 r_i r_j r_k + B_1 (r_i \delta_{jk} + r_j \delta_{ik}) + D_1 r_k \delta_{ij} 
\nonumber \, , \\
S^2_{ijk} ({\bf r}) &=& \langle b_i ({\bf x}) b_j ({\bf x}) v_k ({\bf x}^{\prime}) \rangle \\
&=& A_2 r_i r_j r_k + B_2 (r_i \delta_{jk} + r_j \delta_{ik}) + D_2 r_k \delta_{ij} 
\nonumber \, , \\
S^3_{ijk} ({\bf r}) &=& \langle v_i ({\bf x}) b_j ({\bf x}) b_k ({\bf x}^{\prime}) \rangle \\
&=& A_3 r_i r_j r_k + B_3 r_i \delta_{jk} + C_3 r_j \delta_{ik} + D_3 r_k \delta_{ij} 
\nonumber \, , \\
S^4_{ijk} ({\bf r}) &=& \langle J_i ({\bf x}) b_j ({\bf x}) b_k ({\bf x}^{\prime}) \rangle \\
&=& A_4 r_i r_j r_k + B_4 r_i \delta_{jk} + C_4 r_j \delta_{ik} + D_4 r_k \delta_{ij} 
\nonumber \, ,
\end{eqnarray}
where $A_m$, $B_m$, $C_m$ and $D_m$ are arbitrary functions of $r^2$. Note that the last two 
tensors are {\it not} symmetric in the suffixes $i$ and $j$ which makes an important difference with 
Navier-Stokes fluids where only the velocity field is used to build the third-order correlation tensor. 
The direct consequence is that we need not three but four arbitrary functions to define initially these 
tensors. In the same way as before, we use the continuity condition, $\partial_{r_k} S^m_{ijk} ({\bf r})=0$, 
to constrain our system (Batchelor, 1953); it leads to the following relations
\begin{eqnarray}
r \partial_r A_{1,2} + 5A_{1,2} + 2r^{-1} \partial_r B_{1,2} &=& 0 \, , \label{c1} \\
r \partial_r D_{1,2} + 3D_{1,2} + 2B_{1,2} &=& 0 \, , \label{c2} \\
r \partial_r A_{3,4} + 5A_{3,4} + r^{-1} \partial_r B_{3,4} + r^{-1} \partial_r C_{3,4} &=& 0 \, , \label{c4} \\
r \partial_r D_{3,4} + 3D_{3,4} + B_{3,4} + C_{3,4} &=& 0 \, . \label{c5}
\end{eqnarray}
Additionally, we note that $S^m_{iik} ({\bf r})=0$ whatever the value of $m$ is (since it is a solenoidal
first-order isotropic tensor);
it gives the relations 
\begin{eqnarray}
A_{1,2}r^2 + 2B_{1,2} + 3D_{1,2} &=& 0 \, , \label{c3a} \\
A_{3,4} r^2 + B_{3,4} + C_{3,4} + 3D_{3,4} &=& 0 \, . \label{c3b}
\end{eqnarray}
We introduce now the basic functions which include the parallel and perpendicular components of the 
fields. We have 
\begin{eqnarray}
S^{1,2}_{\parallel \parallel \parallel} &=& A_{1,2} r^3 + (2B_{1,2}+D_{1,2}) r =  Y_{1,2} K_{1,2}(r) \, , \\
S^{1,2}_{\perp \perp \parallel} &=& D_{1,2} r = Y_{1,2} h_{1,2}(r) \, , \\
S^{1,2}_{\parallel \perp \perp} &=& B_{1,2} r = Y_{1,2} q_{1,2}(r) \, , \\
S^{3,4}_{\parallel \parallel \parallel} &=& A_{3,4} r^3 + (B_{3,4}+C_{3,4}+D_{3,4}) r =  Y_{3,4} K_{3,4}(r) \, , 
\hspace{.6cm}
\\
S^{3,4}_{\perp \perp \parallel} &=& D_{3,4} r = Y_{3,4} h_{3,4}(r) \, , \\
S^{3,4}_{\parallel \perp \perp} &=& B_{3,4} r = Y_{3,4} q_{3,4}(r) \, , \\
S^{3,4}_{\perp \parallel \perp} &=& C_{3,4} r = Y_{3,4} s_{3,4}(r) \, ,
\end{eqnarray}
where $K_m$, $h_m$, $q_m$ and $s_m$ are odd scalar functions and $Y_1=v^3$, $Y_2=Y_3=vb^2$, 
$Y_4=Jb^2$. Conditions (\ref{c1})--(\ref{c5}) and (\ref{c3a})--(\ref{c3b}) simplify the expression of the 
third-order tensors, which write finally as
\be
S^{1,2}_{ijk} ({\bf r}) = Y_{1,2} \left[ \left({K_{1,2}-r\partial_r K_{1,2}} \over {2r^3}\right) r_i r_j r_k \right.
\label{sijk1}
\ee
$$
\left. + \left({2K_{1,2}+r \partial_r K_{1,2}} \over {4r}\right) (r_i \delta_{jk} + r_j \delta_{ik}) - 
{K_{1,2} \over 2r} r_k \delta_{ij} \right] \, ,$$
\be
S^{3,4}_{ijk} ({\bf r}) = Y_{3,4} \left[ \left({K_{3,4}-r\partial_r K_{3,4}} \over {2r^3}\right) r_i r_j r_k 
+ {q_{3,4} \over r} r_i \delta_{jk}
\right.
\label{sijk2}
\ee
$$
\left. + \left({K_{3,4}+r \partial_r K_{3,4}/2 -q_{3,4}} \over {r}\right) r_j \delta_{ik} - 
{K_{3,4} \over 2r} r_k \delta_{ij} \right] \, .$$

The first main goal of this Letter is the derivation of the vKH equations for 3D Hall MHD fluids. We start 
with equations (\ref{E1})--(\ref{E2}) and use previous relations derived from the kinematics to find
\begin{eqnarray}
\partial_t R^v_{ij} ({\bf r}) &=& 
\langle v_i \partial_t v_j^{\prime} \rangle + \langle v_j^{\prime} \partial_t v_i \rangle \\
&=& \langle v_i b^{\prime}_\ell \partial^{\prime}_\ell b_j^{\prime} \rangle -
\langle v_i v^{\prime}_\ell \partial^{\prime}_\ell v_j^{\prime} \rangle - 
\langle v_i \partial^{\prime}_j P^{\prime} \rangle + 
\langle v^{\prime}_j b_\ell \partial_\ell b_i \rangle \nonumber \\
&-& \langle v^{\prime}_j v_\ell \partial_\ell v_i \rangle - \langle v^{\prime}_j \partial_i P \rangle 
+ \nu \langle v_i {\partial^{\prime}}^2_{\ell \ell} v^{\prime}_j \rangle
+ \nu \langle v^{\prime}_j \partial^2_{\ell \ell} v_i \rangle \nonumber \, , 
\end{eqnarray}
\begin{eqnarray}
\partial_t R^b_{ij} ({\bf r}) &=& 
\langle b_i \partial_t b_j^{\prime} \rangle + \langle b_j^{\prime} \partial_t b_i \rangle \\
&=& \langle b_i b^{\prime}_\ell \partial^{\prime}_\ell v_j^{\prime} \rangle -
\langle b_i v^{\prime}_\ell \partial^{\prime}_\ell b_j^{\prime} \rangle + 
\langle b^{\prime}_j b_\ell \partial_\ell v_i \rangle
- \langle b^{\prime}_j v_\ell \partial_\ell b_i \rangle \nonumber \\
&+& d_I (\langle b_i J^{\prime}_\ell \partial^{\prime}_\ell b_j^{\prime} \rangle
- \langle b_i b^{\prime}_\ell \partial^{\prime}_\ell J_j^{\prime} \rangle +
\langle b^{\prime}_j J_\ell \partial_\ell b_i \rangle \nonumber \\
&-& \langle b^{\prime}_j b_\ell \partial_\ell J_i \rangle)
+ \eta \langle b_i {\partial^{\prime}}^2_{\ell \ell} b^{\prime}_j \rangle
+ \eta \langle b^{\prime}_j \partial^2_{\ell \ell} b_i \rangle \nonumber \, . 
\end{eqnarray}
After simple manipulations where we use, in particular, the divergence free condition and the 
homogeneity assumption, we get 
\begin{eqnarray}
\partial_t R^v_{ij} &=& \partial_{r_\ell} \left( 
S^1_{i \ell j } + S^1_{j \ell i} - S^2_{i \ell j } - S^2_{j \ell i} \right) 
+ 2 \nu \partial^2_{r_\ell r_\ell} R^v_{ij} , \, \, \, \, \, \, \, \, \, \\
\partial_t R^b_{ij} &=& \partial_{r_\ell} \left( S^3_{\ell j i} - S^3_{j \ell i} + S^3_{\ell i j} 
- S^3_{i \ell j} + d_I S^4_{j \ell i} \right.\\
&-& \left. d_IS^4_{\ell j i} + d_IS^4_{i \ell j} - d_IS^4_{\ell i j} \right) + 2 \eta \partial^2_{r_\ell r_\ell} 
R^b_{ij} \nonumber \, .
\end{eqnarray}
Note that the pressure terms are suppressed because of isotropy (Batchelor, 1953). These general 
dynamical equations reduce to a simple form for the diagonal part of the energy tensor, 
\begin{eqnarray}
\partial_t R^v_{ii} &=& 2 \partial_{r_\ell} \left( S^1_{i \ell i} - S^2_{i \ell i} \right) 
+ 2 \nu \partial^2_{r_\ell r_\ell} R^v_{ii} \, , \label{rii1} \\
\partial_t R^b_{ii} &=& 2 \partial_{r_\ell} \left( S^3_{\ell i i} - S^3_{i \ell i} + d_I (S^4_{i \ell i} 
- S^4_{\ell i i}) \right) + 2 \eta \partial^2_{r_\ell r_\ell} R^b_{ii} . \hspace{.6cm} \label{rii2}
\end{eqnarray}
It is the basic equations from which it will be possible to derive the equivalent of the vKH relations. 
The introduction of (\ref{sijk1}) and (\ref{sijk2}) into (\ref{rii1}) and (\ref{rii2}) gives 
\begin{eqnarray}
\partial_t R^v_{ii} &=& \partial_{r_\ell} \left( {v^3 \over r} r_\ell (4 + r \partial_r ) K_1 
- {vb^2 \over r} r_\ell (4 + r \partial_r ) K_2 \right) \nonumber \\
&+& 2 \nu \partial^2_{r_\ell r_\ell} R^v_{ii} \, , \label{rii2a} \\
\partial_t R^b_{ii} &=& \partial_{r_\ell} \left( - {4 vb^2 \over r} r_\ell (K_3 + r\partial_r K_3 /2 - 2q_3) 
\right. \\ 
&+& \left. {4 d_IJb^2 \over r} r_\ell \left( K_4 + r \partial_r K_4 / 2 -2q_4 \right) \right) 
+ 2 \eta \partial^2_{r_\ell r_\ell} R^b_{ii} \, . \nonumber \label{rii2b}
\end{eqnarray}
By introducing
\begin{eqnarray}
{\tilde K}_m = {1 \over r^4} \partial_r (r^4 K_m)   \, , \, \, 
{\tilde K}_n = {K_n + r \partial_r K_n / 2 - 2q_n \over r} \, , \, \, 
\end{eqnarray}
for $m=(1,2)$ and $n= (3,4)$, we obtain
\begin{eqnarray}
\partial_t \left( (3 + r \partial_r) f^v v^2 \right) &=& v^3 \partial_{r_\ell} ( r_\ell {\tilde K}_1 ) 
- vb^2  \partial_{r_\ell} ( r_\ell {\tilde K}_2 ) \nonumber \\
&+& 2 \nu \partial^2_{r_\ell r_\ell} \left( (3 + r \partial_r) f^v v^2 \right) \, , \label{rii3a} \\
\partial_t \left( (3 + r \partial_r) f^b b^2 \right) &=& 
4d_IJb^2 \partial_{r_\ell} (r_\ell {\tilde K}_4 ) - 4vb^2 \partial_{r_\ell} (r_\ell {\tilde K}_3 ) \nonumber \\
&+& 2 \eta \partial^2_{r_\ell r_\ell} \left( (3 + r \partial_r) f^b b^2 \right) \, .
\label{rii3b}
\end{eqnarray}
By noting the following identity (for isotropic turbulence) 
$\partial^2_{r_\ell r_\ell} =  \partial^2_{rr} + (2 / r) \partial_r$, and the more subtle relation
\be
( \partial^2_{rr} + {2 \over r} \partial_r) (3 + r \partial_r) = 
(3 + r \partial_r) {1 \over r^4} \partial_r (r^4 \partial_r) \, , 
\ee
we finally obtain after some simple manipulations
\begin{eqnarray}
\partial_t \left( (3 + r \partial_r) f^v v^2 \right) = v^3 (3 + r \partial_r) {\tilde K}_1 \\
- vb^2 (3 + r \partial_r) {\tilde K}_2 + 2 \nu \left( 3 + r \partial_r \right) {1 \over r^4} 
\partial_r (r^4 \partial_r f^v v^2) \, , \nonumber \label{rii4a} \\
\partial_t \left( (3 + r \partial_r) f^b b^2 \right) = 4d_IJb^2 \left( 3 + r \partial_r \right) {\tilde K}_4 \\
- 4vb^2 \left( 3 + r \partial_r \right) {\tilde K}_3 + 2 \eta \left( 3 + r \partial_r \right) {1 \over r^4} 
\partial_r (r^4 \partial_r f^b b^2) \, . \nonumber \label{rii4b}
\end{eqnarray}
A first integral of these equations is 
\begin{eqnarray}
\partial_t f^v v^2 &=& v^3 {\tilde K}_1 - vb^2 {\tilde K}_2 + {2\nu \over r^4} \partial_r (r^4 \partial_r f^v v^2) 
\, , \label{rii5a} \\
\partial_t f^b b^2 &=& - 4vb^2 {\tilde K}_3 + 4d_IJb^2 {\tilde K}_4 + 
{2\eta \over r^4} \partial_r (r^4 \partial_r f^b b^2)  \, .  \hspace{.6cm} \label{rii5b}
\end{eqnarray}
These exact equations are the vKH relations for 3D Hall MHD. It is the first main result of this paper.

We shall derive now the equivalent of the four-fifths law found by Kolmogorov for Navier-Stokes 
fluids (Kolmogorov, 1941). We note the relation 
$R^X_{\parallel \parallel} ({\bf r}) = 
\langle X^2_{\parallel} \rangle - {1 \over 2} B^X_{\parallel \parallel} ({\bf r})$, 
with the general form for the structure function
\begin{eqnarray}
B^X_{i j} ({\bf r}) &=& 
\langle (X_i({\bf x}^{\prime}) - X_i({\bf x}))(X_j({\bf x}^{\prime}) - X_j({\bf x})) \rangle \, ,
\end{eqnarray}
where $X=(v,b)$. Introducing this relation into the vKH equations (\ref{rii5a})--(\ref{rii5b}), we get
\begin{eqnarray}
\partial_t \langle v^2_{\parallel} \rangle - {1 \over 2} \partial_t B^v_{\parallel \parallel} &=& 
v^3 {\tilde K}_1 - vb^2 {\tilde K}_2 \\
&+& {2\nu \over r^4} \partial_r (r^4 \partial_r (\langle v^2_{\parallel} \rangle - 
{1 \over 2} B^v_{\parallel \parallel}))  \, , \nonumber \label{rii6a} \\
\partial_t \langle b^2_{\parallel} \rangle - {1 \over 2} \partial_t B^b_{\parallel \parallel} &=&
4d_IJb^2 {\tilde K}_4 - 4vb^2 {\tilde K}_3 \\
&+& {2\eta \over r^4} \partial_r (r^4 \partial_r (\langle b^2_{\parallel} \rangle - 
{1 \over 2} B^b_{\parallel \parallel})) \, , \nonumber \label{rii6b}
\end{eqnarray}
and eventually
\begin{eqnarray}
\partial_t \langle v^2_{\parallel} \rangle - \partial_t {B^v_{\parallel \parallel} \over 2} &=& 
v^3 {\tilde K}_1 - vb^2 {\tilde K}_2 \label{rii7a} 
- {\nu \over r^4} \partial_r (r^4 \partial_r B^v_{\parallel \parallel}) , \hspace{.6cm} \\
\partial_t \langle b^2_{\parallel} \rangle - \partial_t {B^b_{\parallel \parallel} \over 2} &=&
4d_IJb^2 {\tilde K}_4 - 4vb^2 {\tilde K}_3 \label{rii7b} 
- {\eta \over r^4} \partial_r (r^4 \partial_r B^b_{\parallel \parallel}) . \, \, \, \, \, \, \, \, \, \, \label{rii7b} 
\end{eqnarray}

We define the mean (total) energy dissipation rate per unit mass, $\varepsilon^T$, for isotropic 
turbulence, as
\be
\partial_t \langle v_{\parallel}^2 + b_{\parallel}^2 \rangle = - (2/3) \varepsilon^T \, . 
\label{rate}
\ee
Exact scaling laws for third-order correlation tensors may be derived from the previous relations 
(\ref{rii7a})--(\ref{rii7b}) by 
assuming the following assumptions specific to fully developed turbulence (Frisch, 1995). We first 
consider the long time limit for which a stationary state is reached with a finite $\varepsilon^T$. 
Second, we take the infinite (magnetic) Reynolds number limit ($\nu \to 0$ and $\eta \to 0$) for which 
the mean energy dissipation rate per unit mass tends to a finite positive limit. Therefore, in the inertial 
range, we obtain at first order the relations 
\begin{eqnarray}
- {1 \over 6} \varepsilon^T r &=& v^3 (K_1 + r\partial_r K_1/4) - vb^2 (K_2 + r\partial_r K_2/4) \nonumber \\
&-& vb^2 (K_3 + r \partial_r K_3 / 2 - 2q_3) \\
&+& d_I Jb^2 (K_4 + r \partial_r K_4 / 2 - 2q_4) \nonumber \, , 
\end{eqnarray}
which can also be written as
\begin{eqnarray}
- {1 \over 6} \varepsilon^T r &=& 
(S^1_{\parallel \perp \perp} + {1 \over 2} S^1_{\parallel \parallel \parallel} )
- (S^2_{\parallel \perp \perp} + {1 \over 2} S^2_{\parallel \parallel \parallel})  \\
&+& (S^3_{\parallel \perp \perp} -S^3_{\perp \parallel \perp})
- d_I (S^4_{\parallel \perp \perp} - S^4_{\perp \parallel \perp}) \, . \nonumber \label{law2b}
\end{eqnarray}
The last step consists in introducing structure functions which gives, after some manipulations, the 
final result
\begin{eqnarray}
- {4 \over 3} \varepsilon^T r &=& 
B^{vvv}_{\parallel i i} + B^{vbb}_{\parallel i i} - 2 B^{bvb}_{\parallel i i} 
- 4 d_I (S^4_{\parallel i i } - S^4_{i \parallel i}) , \hspace{.5cm} \label{law2c} \\
&=& B^{vvv}_{\parallel i i} + B^{vbb}_{\parallel i i} - 2 B^{bvb}_{\parallel i i} 
+ 4 d_I \langle [({\bf J} \times {\bf b}) \times {\bf b}^{\prime}]_{\parallel} \rangle \, , \nonumber \label{law2c}
\end{eqnarray}
with $B^{\alpha \beta \gamma}_{i j k} = 
\langle (\alpha_i^{\prime} - \alpha_i)(\beta_j^{\prime} - \beta_j) (\gamma_k^{\prime} - \gamma_k) \rangle$.

Equations (\ref{rii5a})--(\ref{rii5b}) and (\ref{law2b}) are the main results of this paper. 
The former equations are exact for homogeneous and isotropic turbulence and the latter assumed 
additionally the existence of a large inertial range on which the total energy flux is finite 
and constant. The most remarkable aspect of these laws is that they do not only provide a 
linear scaling for the third-order correlation tensors within the inertial range of length scales, but they 
also fix the value of the numerical factor appearing in front of the scaling relations. 
Another important remark is about the fields used to build the third-order correlation tensors. Indeed, 
the convenient variables are not only the velocity and magnetic field components but also the current 
density components. Note that attempts to find a simple expression in terms of only structure functions 
failed, and therefore relations (\ref{law2b}) seem to be the most appropriate. A similar situation was 
found for example in MHD flows when the magnetic helicity is included in the analysis (Politano and 
Pouquet, 1998; Politano, Gomez and Pouquet, 2003). 

The vKH equations (\ref{rii5a})--(\ref{rii5b}) derived here in the framework of Hall MHD are compatible 
with the one derived by Chandrasekhar (1950) for MHD when the large-scale limit ($d_I\to 0$) 
is taken. Note that some minor manipulations have to be made in (Chandrasekhar, 1950) to prove the 
compatibility since the notation are not the same (for example, we have $P \equiv vb^2 {\tilde K}_3$). 
As explained above, the notation used here seems to be more suitable for Hall MHD which therefore 
requires for a better understanding a complete re-derivation of the dynamical equations. In the same way, 
when the large-scale limit is taken, relation (\ref{law2b}) is compatible with previous works (Politano and 
Pouquet, 1998; Politano, Gomez and Pouquet, 2003), which are also compatible with Navier-Stokes fluids 
when additionally the magnetic field is taken equal to zero. 

The exact results found here provide a better theoretical understanding of Hall MHD flows. They show 
that the scaling relation does not change its power dependence in the separation $r$ at small-scales if 
the statistical correlation tensor used is modified. The interesting point to note is the compatibility with 
previous heuristic and numerical results (Biskamp et al., 1996). Indeed, a simple dimensional analysis gives
the relations $r \sim b^3$ for large-scales, and $r^2 \sim b^3$ for small-scales (since $J \sim b/r$), 
which give respectively the magnetic energy spectrum $E \sim k^{-5/3}$ and $E \sim k^{-7/3}$. 
Therefore and contrary to the appearance, the exact results found may provide a double scaling relation. 

These multi-scale laws provide a relevant tool to investigate the non-linear nature of the high frequency 
magnetic field fluctuations in the solar wind whose (dissipative {\it vs} dispersive) origin is still controversial 
(Goldstein et al., 1994; Markovskii et al., 2006; Stawicki et al., 2001; Galtier, 2006ab; Galtier and Buchlin, 
2007). The use of multi-point data may give information about both the magnetic field 
and the current density which can be used to check the theoretical scaling relations. 
The observation of such a scaling law would be an additional evidence for the presence of a dispersive 
inertial range and therefore for the turbulent nature of the high frequency magnetic field fluctuations. 
The recent observation of the Yaglom MHD scaling law (Sorriso-valvo et al., 2007) at low frequency 
provides a direct 
evidence for the presence of an inertial energy cascade in the solar wind. The theoretical results given 
here allows now to extend this type of analysis to the high frequency magnetic field fluctuations and, 
more generally speaking, to better understand the role of the Hall effect in astrophysics, like \eg for the 
magnetorotational instability in cool proto-stellar disks (Wardle, 1999; Balbus and Terquem, 2001), or 
in laboratory fusion plasmas. 
In such situations, isotropy is often broken because of the presence of a strong large-scale magnetic 
field (see \eg Muller et al., 2003) and a generalization of the present description to anisotropic turbulence is 
then necessary.

Financial support from PNST/INSU/CNRS are gratefully acknowledged. I would like to thank 
H. Politano for useful discussion. 

\end{document}